# GALILEAN SATELLITES AS SITES FOR INCIPIENT LIFE, AND THE EARTH AS ITS SHELTER

E.M.Drobyshevski

*Ioffe Physico-Technical Institute, Russian Academy of Sciences,194021 St.Petersburg*
*E-mail: emdrob@mail.ioffe.ru*

ABSTRACT. It appears fairly obvious that life could evolve only in the presence of water. However, from the outset the Earth and Mars had very little water (~0.03%) and protoorganic substances (several orders of magnitude less than that), which poses two problems, namely,
(1) a concentration gap, where a reasonable original concentration of aminoacids in water could not be higher than ~$10^{-8}$ kg/l, which made their fusion to more complex formations unlikely, and
(2) inhibition of this fusion by salts like NaCl, $MgSO_4$ etc., whose high concentration is due to their being washed out of rocks by small amounts of water.

An opposite situation occurred on the Galilean satellites, where water constituted ~50%, hydrocarbons ~5-10%, and the nitrate-based organics ~1% (the estimates are based on the composition of cometary nuclei). Liquid water containing only minute amounts of salts could exist here both after the satellite accretion and during hundreds of Myrs following regular global explosions of the icy envelope saturated by $2H_2+O_2$, i.e., the products of volumetric electrolysis of these ices due to the action of the electric currents excited by the strong (~$10^2$ Gauss) ancient magnetic field of Jupiter. On Io, two to three such explosions occurred altogether, on Europa, two, and on Ganymede, only one. The ices of Callisto did not explode. It appears that Titan also suffered an explosion, only ~10 kyr ago, a conjecture corroborated by the fact that many of our predictions have obtained observational support.

An explosion ejects numerous fragments of the outer icy-envelope layers, likewise saturated by organics and electrolysis products. These are the nuclei of the SP comets, whose behavior and chemical manifestations are accounted for by such a composition.

An explosion creates on the satellites a deep (700-800 km) hot ocean of a gradually cooling almost fresh water. Part of the pyrolized organics ascends to the surface, while the other, the heavier one, settles down to the bottom. Some of the ice fragments of the primary composition ejected in an explosion continue to fall into the ocean during a long time, including the period after its cooling and formation of ice on its surface.

In these conditions, the problems of salt excess or a concentration gap simply do not exist (the concentration of organic substances near the surface of the ocean and on its bottom may reach as high as 100%). The thermodynamic parameters of volatiles vary in the ranges inconceivable for the Earth, namely, $T \sim 10^2$–$10^3$ K, $p \sim 0$–$10^4$ atm, with the pristine and highly pyrolized components being in direct contact. Powerful energy flows, including the electrical ones, accompanied by intensive convection motions set in. One may conjecture the formation of a rich variety of catalytic substances. Absolute enantiomeric synthesis could take place in electrolysis in the presence of a magnetic field. Subsequent explosions of icy envelopes make the conditions still more diverse. It thus appears that the probability of synthesis of biological compounds here is much higher than that on the Earth.

The Earth, because of its milder conditions, could provide an ideal habitat for development of the ready-to-evolve primitive life forms. It thus becomes clear that the assumption of ejection of cometary nuclei from icy moon-like bodies removes the problem of transfer of the life borne on them to the Earth. Interestingly, the multiplicity of envelope explosions on Io and Europa, where, thus, protobiological objects could form and evolve in repeatedly occurring extremal conditions, may shed light on the nature of intermittent stages of the development of biota on the Earth, of the type of the pre-Cambrian extinction followed by subsequent explosive growth of life in a new direction etc.

**Introduction. The difficulties created by the assumption of the life having originated on the Earth**

That life exists on the Earth, although its rigorous definition is still lacking, and we have to be content with quasi-intuitive notions, cannot apparently be questioned. Where else life exists, had been, or may form, is already unclear. How it appeared on the Earth - is likewise an unsolved issue. It appears desirable, and present-day science puts forward such claims, to find an answer to this problem from the standpoint of positive knowledge. One could, of course, invoke the concept of panspermia, but, in view of the finite age of the Universe, this does not remove the fundamental question of the origin of life in general.

It is only natural to start with the assumption that life was borne on the Earth. This assumption meets, however, immediately with difficulties of at least two kinds: first, the conditions on the Earth are comparatively mild and monotonous in the sense that they are favorable for the existence of already formed organisms but not for abiogenic synthesis of their constituent molecules and radicals. The very first researchers were already well aware of this point. This is why Miller (1953) and his followers used prolonged electric discharge in gases of a specially chosen composition (a mixture of $CH_4$, $NH_3$, $H_2O$, $H_2$), which, as they believed, could simulate the primeval atmosphere. Attempts were also undertaken to study the extremal conditions in volcanoes (e.g. Washington, 2000), etc.

Here we encounter, however, another problem, namely, that of a concentration gap (Folsom, 1979). Indeed, although thunder does roar in the skies, as it seems sometimes, continuously, the concentrations of abiogenic organics produced by the lightning are so low as to be incomparable with the experiments of Miller. The organic molecules thus produced cannot practically interact with one another and are doomed to destruction by a variety of factors. Among them is, for instance, UV irradiation; however, the most important factor is the chemical activity of the minerals dissolved in water, primarily, of the salts of the kind of $NaCl$, $MgSO_4$, and the like.

Their concentration in water on the Earth reaches very high levels (~30 g/l), because it is due to the salts being washed out of large masses of rock by small amounts of water ($\leq$ 0.03%). By combining with the active centers of the as-formed organic molecules and radicals, the $Na^+$, $Mg^{++}$, $Cl^-$ etc. ions deactivate them, which makes the molecules incapable of further interactions (Folsom, 1979).

The reason for which all compounds of biogenic origin have laevo-chirality (provided naturally that the molecules of a given compound can be chiral) is also unknown. Products obtained by abiogenic laboratory synthesis (without a bioseed) demonstrate an equal probability (racemic process) for the molecules of both types to form. From the times of L.Pasteur (e.g. Avetisov and Goldanski, 1996; Bonner, 1996, and refs. therein), many attempts were made to explain this preferential chirality, but none of them proved successful enough. For instance, the recent studies of photochemical processes in a rarefied interstellar medium (Bernstein et al., 2002; Munoz Caro *et al*., 2002; Rikken and Raupach, 2000) also meet with the problem of the concentration gap as soon as the discussion reaches the point of the consequences of the fallout of their products on the Earth. The assumption of the initial prevalence of L-chiral molecules having been of a purely random nature appears to be most consistent (Avetisov and Goldanski, 1996). However, it hardly can provide a feeling of satisfaction; indeed, L-chiral organisms are not capable of consuming the D-chiral ones, i.e., they cannot be their active competitors, which weakens the action of natural selection in the observed direction. Two worlds residing, possibly, in different stages of development, could (and should) have been coexisting simultaneously without interference. But it just did not happen! One could conceive here a certain analogy with the practical absence of antimatter in our Universe. It may be conjectured that it is the latter factor that accounts for the L-chirality in our present life. One should only find the specific physical relations.

We are going to show that all the above problems are removed within the Solar System, if one considers the possibility of abiogenic synthesis of organics in an ensemble of icy moonlike planetary bodies of the type of the Galilean satellites. Due to the very strong Jupiter's magnetic field (up to $\sim 10^3$ Gauss in the past), the conditions for a manifestation of chemical activity reach here extreme values and a diversity inconceivable on the Earth. Note that these conclusions tie in closely with the data obtained in the recent decades on the chemical manifestations, the structure and origin of the comets and other minor bodies (Drobyshevski *et al*., 1995), as well as on the specific features of, and individual differences between, the Galilean satellites (Drobyshevski, 1979, 1980) and Titan (Drobyshevski, 1981, 2000) , and of the Martian satellites and the irregular satellites of Jupiter.

## 2. Conditions on the Galilean satellites. The possibility and consequences of explosion and melting of their ices

Condensed water in the Solar system can be found not only on the Earth. It is present as ice on Mars, probably in the near-polar regions of the Moon, and makes up a large part of the material of such bodies as the moonlike Ganymede, Callisto, Titan, Triton, Pluto, many small satellites (like Deimos etc.), and of the comets, which are actually fragments of the outermost layers of moonlike bodies (as we shall see below, they can be shed off by two different mechanisms, more specifically, by an explosion or an impact). The objects making up the hypothetical Oort's cloud or Kuiper's cometary belt also should consist primarily of ice. However, as follows from the cosmogony dealing with planetary systems, in particular, the Jupiter−Sun system together with close binary stars (Drobyshevski, 1974), there should exist in the near periphery of the Solar system a joint planetary-cometary cloud consisting of scores of icy Pluto-like planets and icy fragments produced in their extremely infrequent collisions, i.e., nuclei of long-period (LP) comets (Drobyshevski, 1978).

Liquid water is needed for the creation and support of life. Besides the Earth, it is possibly present under the thick icy crust of Europa (the total ocean thickness ~100 km) (Squyres *et al.*, 1983). It does not freeze out here because of the tidal energy release, which accounts for the active volcanism on Io, a closer satellite of Jupiter. Ancient traces of water flows can be seen on Mars. These observations provide a stimulus to studies aimed at a search for signs of life (even if already extinct) on Mars and Europa.

On the other hand, one should not overlook the fact that all icy moonlike bodies could have possessed liquid envelopes in the course of accretion and after it has come to an end. For a thickness of ~700−800 km, the time taken by their subsequent freezing could be as long as ~ 100 Myr and longer. Moreover, the above cosmogony treating the Jupiter−Sun system as a limiting case of a close binary suggests that all Galilean satellites were initially very close in composition (Drobyshevski, 1980), i.e., that they contained up to 50−60% of ices of volatile compounds (water, and up to ~10% of carbon and nitrogen compounds, including primitive abiogenic organics etc.).

It should be pointed out that the naive scenario of ice removal to the periphery of the satellite system in the course of satellite accretion due to heating by radiation emitted by the hot contracting Jupiter is not supported by any calculations; in fact, all of them suggest that the temperature in the hypothetical protosatellite disk in the orbit of Io was <150 K (Bodenheimer *et al.*, 1980).

An analysis of possible reasons for the compositional differences and other features on the Galilean satellites brought us to the conclusion that they originate from the action of the very strong magnetic field of Jupiter (see below). All the known differences between the satellites can be accounted for by repeated global explosions of their massive icy envelopes saturated by the solid solution of $2H_2+O_2$ in ice, i.e., by products of volumetric electrolysis of dirty ice (Drobyshevski, 1980; Drobyshevski et al., 1995).

The electric current required for the electrolysis is generated through the motion of the satellite in the planet's magnetosphere. Direct measurements made by Voyager-1 yielded ~ 4.8 MA for the current flowing presently through one wing of the magnetic tube enclosing Io (Ness *et al*., 1979a). Indications of a strong magnetic field perturbation caused by the generated currents were observed near Ganymede (Ness *et al*., 1979b).

The electrical conductivity of ice is due to the proton mobility (Decroly *et al*., 1957; Petrenko and Chesnakov, 1990) and increases, as a rule, by many orders of magnitude in the presence of impurities (Hobbs, 1974; Petrenko and Whitworth, 1999), a situation characteristic of the icy envelopes of moonlike bodies. The conduction of metal oxides, carbonaceous, silicate and similar inclusions in ice is of the electronic (or hole) character. Therefore current flow through the inclusion interfaces results in electrolytic ice decomposition. Ice electrolysis is a process well known for a long time (Decroly *et al*., 1957; Petrenko and Whitworth, 1999). Because of the high pressure in the envelopes (~ $10^3$–$10^4$ atm), the electrolysis products, $2H_2+O_2$, rather than evolving in the form of gases, build up in the ice in the form of a solid solution (clathrate hydrates) and become redistributed throughout the envelope volume by solid-state thermal convection. For contents $2H_2+O_2$ ~ 10–15 wt %, ice becomes capable of detonation (Drobyshevski, 1986). A strong enough meteorite impact initiates off-central explosion of the icy envelope. Bodies with masses <$1M_{moon}$ lose practically all of their envelope ices (the case of the Pluto-like Phaeton between Mars and Jupiter (Drobyshevski *et al.*, 1994)), while for $M > 1.5M_{moon}$, the loss is only a few tens of percent of the envelope mass. The lost material consists of the vapors of water and other volatiles, rock inclusions in ice, and km-sized unexploded ice fragments of the outermost cold envelope layers. The latter also contain electrolysis products, so that under certain conditions they are capable of new explosions or combustion. These are nuclei of typical SP comets.

Their manifestations (jet emissions of a dusty gas with velocities ~1 km/s, distant bursts and breakups into fragments separating with velocities ~1–10 m/s and so on) and chemical activity (the presence of all kinds of hot ions, including heavy organic radicals in the immediate vicinity of the nucleus, submicron-sized CHON dust etc.) find readily interpretation within the explosion paradigm by combustion of sublimation products at a lack of oxygen (Drobyshevski, 1997), while remaining enigmatic when viewed from the standpoint of the old condensation-sublimation hypotheses. The above reasoning accounts for the differences in the behavior of most SP and LP comets, although bodies can naturally transfer from one group to another by laws of celestial mechanics.

As follows from an analysis of the orbital elements of some LP comets, such an explosion occurred on Titan only ~$10^4$ years ago. This accounts for the formation of its dense atmosphere, the unusually high eccentricity of its orbit, and the formation of young Saturn's rings (Drobyshevski, 1981, 2000). Some of the predictions made by us on this basis ($N_2$, HCN, CO, $CO_2$ in Titan's atmosphere, two, km- and cm-sized, populations in Saturn's rings, the 2585-km radius of Titan's solid surface) have been confirmed, while others (excess heat flux from Titan of ~1.5–15% of the flux received from the Sun, the asynchronous rotation of Titan, and, the most important of all, the existence of a deep liquid ocean underneath a ~ 1-km thick ice crust (Drobyshevski, 1982)) are still awaiting substantiation.

Judging from a number of indications, including the density, topography and the like, explosions of the Galilean satellites occurred repeatedly. Io suffered 2–3 explosions, which culminated in the complete loss of its ices (see, however, below the fate of Amalthio), on Europa two explosions occurred to leave only 8–10% of its total ice mass, Ganymede underwent only one explosion, and on Callisto, none yet (or it occurred so long ago, ~4 Gyr, that all visible traces of it disappeared). If its ices explode now, the Earth will suffer the heaviest bombardment (and subsequent mass extinction) in its history from the time of formation (Drobyshevski, 1989). The explosions on the Galilean satellites and the possibility of subsequent detonation of their icy fragments account very well for the origin and properties of the Trojans and of the irregular satellites of Jupiter.

Estimation of the evolution of Jovian magnetic field required for realization of the above explosion history of the Galilean satellites suggests that the field, rather than being generated now, is most probably of the relic nature and decays exponentially with a characteristic time $\tau \sim$ 0.6–1 Gyr (Drobyshevski, 1980). Similar estimates were obtained also by Stevenson and Ashcroft (1974) for the Ohmic field decay time based on Jupiter stucture models. It thus follows that (*i*) the original magnetic field of Jupiter could have exceeded the present level by a factor $10^2$–$10^3$ and, therefore, (*ii*) the icy envelopes of the satellites were liquid because of the Ohmic heating. The ices of Io solidified completely only ~1.5 Gyr ago, those of Europa, ~2 Gyr, of Ganymede, ~3–4 Gyr, and Callisto's ices could become solid ~4–4.5 Gyr ago. Interestingly, the envelope water could be warm before freezing because of the heating, and on the satellites closest to Jupiter, even hot. On the hypothetical fifth Galilean satellite, Amalthio, which must have lost its ~$1 M_{moon}$ mass as a result of such heating to become the present-day small Amalthea (Drobyshevski, 1979), the water could have reached the boiling point, so that current flow through the ocean could be interrupted for some periods by a thick warm atmosphere with a high water content. The atmosphere was lost not only through thermal dissipation but was also torn off continuously by the *I*×*B* forces acting in its upper ionized layers. The original magnetic field in the orbit of Amalthio could be as high as ~$10^2$ G, which left practically no limit to various manifestations of atmospheric electricity (we may again recall the experiments of Miller). All of the above reasoning about Amalthea may apply to other small satellites inside the Galilean system as well.

Thus, liquid water formed on the Galilean satellites repeatedly and for long periods of time. Deep oceans together with transient atmospheres underwent a rough evolution.

Obviously enough, explosions of the oceans frozen to the bottom, which occurred after solid-state thermal convection had redistributed the impurities and mineral inclusions, as well as the volumetric electrolysis products throughout their volume, introduced a particular variety into the thermodynamics of water and of the prebiological organics contained in the icy envelope, whose presence (~10%, including ~1% nitrogen-containing volatiles) is evidenced by the composition of comets. The temperature increased in the explosions up to ~$10^3$ K, and pressure, to $(3-5)\times10^4$ atm (Drobyshevski, 1986; Drobyshevski *et al.*, 1994). The organic component of the ices underwent both pyrolysis and synthesis of new compounds, which do not form in moderate conditions ($T < 300$ K, $p < 10$ atm). Catalytic action of large amounts of mineral inclusions in the ices could add an additional diversity to the chemical reactions. The fast expansion of the explosion products into vacuum brought about quenching the high-temperature equilibrium constituents and their preservation.

When the ocean condensed back from the products of explosions, the processed organics (both pyrolyzed and newly synthesized) rose to the surface or settled to the bottom, depending on their density and capacity to coagulation, both with itself and with other substances.

Therefore here, in contrast to the Earth, no problem of the concentration gap arises.

Because the mass of water is very large compared with that of the rocks, it cannot contain mineral salts here in so large amounts as on the Earth. The water on the Galilean satellites is, from our viewpoint, practically fresh. The presence of small amounts of inorganic impurities can even add to the diversity of the products of organic synthesis.

An additional factor is introduced by the fall-back on this satellite (and adjacent bodies) of icy fragments having the original, preexplosion chemical composition.

## 3. The role of EM electrolysis. The origin of the preferential chirality of biomolecules

As we have seen, the electrolysis initiated by the interaction of icy moonlike bodies with the magnetic field of Jupiter played a particular part in the evolution of the Galilean satellites (including Amalthio) and in the origin of minor bodies, both comets and asteroids.

Obviously enough, electrochemical processes, especially of such a power that is capable of acting on the global parameters of a planet, should affect substantially the finer processes of synthesis and evolution of organics by creating locally nonequilibrium negentropic situations.

We shall not delve into the details of such processes. Instead, we shall point out only two possible implications.

The first of them relates to the deuterium concentration in the aminoacids being higher than that in water, for instance, in the Murchison meteorite (see, e.g., Kerridge, 1999, and refs. in Bernstein, 2002). Actually, such an enrichment should not come as a surprise in a frame of the electrolysis concept, if one takes into account the noticeable difference between the mobilities of the $H^+$ and $D^+$ ions, which is observed sometimes in solid electrolytes. In various compounds the ratio of their mobilities may vary from the classical value $2^{1/2}$ to ~ $10^2$ (Sherban and Nowick, 1989; Hainovski *et al.*, 1986). The same may concern isotopic shifts of oxygen and similar elements (Drobyshevski, 1995, 2002).

A more substantial implication relates to the synthesis of organic molecules of a preferential chirality in electrolytic processes occurring in the presence of a magnetic field. Already P.Curie suggested a possibility of a magnetic field effect on the activity of products of organic synthesis (De Gennes, 1982). This issue is dealt with in copious theoretical literature (e.g., Avalos *et al.*, 1998; De Gennes, 1982; Rhodes and Dougherty, 1978). All authors agree on the opinion that in conditions close to thermodynamic equilibrium magnetic field cannot exert any influence on the optical activity of the products of chemical reactions. The magnetic field can influence this activity only through the kinetics of the processes involved.

In comparing the thermal energy of a molecule, ~$kT$, with the energy of interaction of its electrical dipole $p$ with an electric field and of its magnetic dipole $\mu_B$ with a magnetic field, Rhodes and Dougherty (1978) and De Gennes (1982) estimate the possible asymmetry $\varepsilon$ between the dextra and laevo reactions (for $\varepsilon \ll 1$) as

$\varepsilon \approx (pE/3kT)(\mu_B B/kT)$.

Assuming $p \approx 1$ eÅ and $\mu_B = 0.927 \times 10^{-20}$ erg/Gauss = 1 Bohr magneton, and taking $E = 10^3$ V/cm and $B = 10^4$ G, Rhodes and Dougherty obtained $\varepsilon \sim 3 \times 10^{-7}$, a figure that did not instill much optimism.

Kinetic processes play a key role in electrolysis. The fields generated in near-electrode double electric layers ~2–3 Å thick can sometimes become as high as $E \approx 10^7$ V/cm. They are high enough to completely orient at $T \sim 300$ K, say, all the water molecules adjacent to the electrode surface (i.e., $pE \geq kT$) (Delahay, 1965; Reeves, 1974). Then for $B = 10^4$ G and $T = 250$ K (water remains liquid at $\approx 2 \times 10^3$ atm) $\mu_B B/kT = 2.7 \times 10^{-3}$. Whence $\varepsilon \approx 10^{-3}$, which exceeds by far the above estimate.

Such estimates are naturally very rough, because they are made without taking many factors into account. The final answer can be obtained only from experiment.

We are aware only of two experiments on electrolysis with a magnetic field applied parallel to an electric field, which were aimed at absolute enantiomeric synthesis, i.e., synthesis carried out in the absence of a catalyst of a given chirality.

The first experiment was performed by Takahashi et al. (1986). The authors reduced the phenylglyoxilic or pyruvic acids in a magnetic field of 980–1,680 G, which was directed normal to the surface of a mercury cathode, to the mandelic or lactic acids. The latter were obtained predominantly in the dextra-rotatory form. The yield of the optically active product was found to be directly proportional to the magnetic field strength and reached 25% for the mandelic acid for 1,680 G. The results obtained did not depend on the field direction. As shown by extrapolation toward lower magnetic fields, an asymmetry in the products of electrolysis becomes noticeable at $B \approx 300$–400 G. On the other hand, for $B > 6,000$–7,000 G one may expect a ~100% yield of the dextra-rotatory mandelic acid. The electrolyte temperature during the experiment was maintained at 10 or 25°C by passing water upward through the jacket surrounding the cell, with the mercury electrode cooled first. In this way a stratification stable against thermal convection was produced in the catholyte. Before the experiment, nitrogen was blown through the electrolyte to free it of oxygen. Mixing the electrolyte during an experiment by bubbling nitrogen or through MHD effects removed the positive result. Unfortunately, the paper is fairly laconic and does not contain a description of many other important details.

Bonner (1990) described in a more detailed paper an attempt at reproducing the above results. However, in fields $B$ = 1,400–70,500 G he managed to obtain racemic products only.

It should be stressed that despite the apparent simplicity, these experiments are very subtle. The influence many parameters can produce remains unclear and, therefore, is poorly controllable.

For instance, the magnetic fields used in the two experiments had noticeably different configurations. Bonner carried out his experiments in considerably stronger fields than Takahashi et al., which could initiate substantial MHD mixing. He did not also take precautionary measures to cool the electrolytic cells, which did not ensure thermal stratification of the catholyte and the mercury cathode that would be stable against convection. Oxygen was not removed from the catholyte before the experiment. The racemic composition of the electrolysis products could be originally stimulated by the use in this experiment of plastic in the cell design, plywood for cell fixtures, and rubber tubes with paper in salt bridges.

To sum up, it appears highly desirable to continue such experiments with magnetic field, which should be performed with utmost care against racemic contamination, electrolyte mixing etc. One should carry out experiments with differently oriented magnetic field gradients, different types of electrodes, and, most importantly, with different electrolytes, i.e., starting organic substances. It is quite probable that in some conditions a practically pure enantiomer of one type would be produced already at $B \sim 10^2$–$10^3$ G.

## 4. Paths of bioproducts exchange among celestial bodies

The predominant chirality of the organic world may be an indication that it was created in nonequilibrium processes in a fairly strong magnetic field ($B \sim 10^2$–$10^3$ G). As we have seen, the conditions necessary for this could be realized on Amalthio. Presently, it is a rocky fragment where life is impossible. What were the paths by which incipient life could be transferred to other bodies?

We have mentioned the tearing off of the outermost ionized part of Amalthio's rough water atmosphere by the $I \times B$ forces. The vapors and snow-flakes could well drag into space both large organic molecules and microscopic organisms. They should not necessarily spend decades to reach the Earth. It was sufficient to get in a few hours to the atmosphere or the surface of one of the neighboring Galilean satellites with largely similar conditions at the time, where the molecules could continue to evolve, but now in a slightly different sense.

Another obvious carrier within this scenario is the showers of cometary nuclei, i.e., numerous fragments of the outermost layers of the electrolyzed icy envelopes of Moon-like bodies shed off in explosions. The bioobjects frozen in large amounts in km-sized ice blocks were already capable of reaching the Earth without being destroyed by cosmic radiation. Ice and rock fragments ejected from planets into space in powerful meteoroid impacts could also act as such carriers (Drobyshevski, 1995).

## 5. Conclusion

Thus, our combined magnetohydrodynamic-electrochemical scenario turns out to be fairly non-contradictory and quite promising.

Starting from only one assumption, well substantiated physically by laboratory-based experiments, of the possibility of volumetric electrolysis of impurity-loaded ices containing mineral inclusions and ~10% primitive organics, one can explain the differences between the Galilean satellites and draw conclusions concerning the relic nature of Jovian magnetic field. This field decays as a result of Ohmic dissipation with a characteristic time $\tau \sim$ 0.6−1.0 Gyr, which correlates well with the independent estimates made by Stevenson and Ashcroft (1974) from Jupiter model calculations. Thus, the ancient Jovian magnetic field exceeded apparently the present level $10^2$−$10^3$ times. In Amalthea's orbit, it could reach ~ 300 G.

One can readily estimate that dissipation of Joule's energy in the Galilean satellites for currents >100 MA was capable of melting their icy envelopes and sustain them in the form of warm oceans for aeons (depending on their distance from Jupiter) (Drobyshevski, 1980). If an icy Galilean satellite were placed in Amalthea's orbit, such an ocean would become hot, and a thick water atmosphere would develop on the satellite. The energetics are such that not only the atmosphere but also the ocean and, thereafter, the rocky core would be scattered into space after a few heating–freezing cycles. The satellite, which we called here Amalthio (Drobyshevski, 1979), would become in a time <1 Gyr a minor rocky remnant, i.e., the present-day Amalthea.

The dramatic events on the Galilean satellites, except, apparently, Callisto, were initiated by global explosions of their icy envelopes, which were saturated as a solid solution by up to 10−15 wt % products of volumetric electrolysis of ices, i.e., $2H_2+O_2$. Explosions entail two consequences: (*i*) a loss of a part of the envelope mass (up to 20−50%) in the form of volatiles and unexploded icy fragments of the outermost layers (which became cometary nuclei), and (*ii*) formation of an atmosphere (of $N_2$ and the lightest products of pyrolyzed organics like $CH_4$, with an admixture of HCN, CO, $CO_2$ and so on, as on Titan) and of a hot ocean of practically fresh water (because the water/rocks ratio on the satellites $\approx 1$, and not ~0.03%, as is the case of the Earth), where the light part of the pyrolized organics floats up to concentrate near the surface, while the heavier one settles down to the bottom. After the subsequent freezing of the ocean, solid-state thermal convection again redistributes these substances (of already a modified composition) and mineral inclusions over the bulk of the icy crust. The ice again becomes "dirty" and, thus, acquires a high electrical conductivity conducive to volumetric electrolysis. The conditions become favorable for new accumulation of the electrolysis products, a new explosion, and formation of a deep warm ocean.

Thus, this explosion scenario does not have two factors which impede the creation of life on the Earth, namely, (*i*) the concentration gap, by which the content of primitive organics in the Earth's water ~10 ppb (whereas here after an explosion the organics concentrate in thick layers in water), and (*ii*) the presence in the Earth's water of large amounts of mineral salts (up to ~3%), whose ions block the active centers of organic molecules and radicals, thus precluding their combining.

The explosions on the Galilean satellites created on a global scale conditions unattainable on the Earth, more specifically, mixtures of primitive carbon- and nitrogen-containing organics with water and finely dispersed mineral substances with temperatures ranging from ~$10^2$ to ~$10^3$ K and pressures, from 0 to ~$10^4$ atm. It should also be added that the satellites received continuously an infall of icy fragments both from the earlier explosions and from those on neighboring bodies, which carried in a frozen form both the primordial organics and the organic mixtures that had already passed through some evolutionary cycles on this or other bodies. The in-falling material contains also quenched products of explosion-generated high-temperature equilibrium (which provides the simplest possible explanation, for instance, for the presence of HCN in Titan's atmosphere). One might say that, considered from the standpoint of early biological evolution, the Galilean satellite system behaved as a unique, closely interacting ensemble.

The electrochemical processes occurring at global currents of $\geq 10^8$ A acted as a powerful factor, which could hardly play a noticeable role on the Earth. They were capable not only of decomposition of water or ice into $2H_2+O_2$ but of synthesis of a rich variety of organic and inorganic compounds in diverse combinations as well. Electrochemical processes are actually thermodynamically nonequilibrium transformations. It appears therefore tempting to assume that application of a magnetic field of $10^2$–$10^3$ G could initiate here synthesis of optically active organic compounds. Unfortunately, perfect confidence in the realizability of this possibility is lacking. Indeed, one could not conceive until very recently that fields of such strength could bear any relation to the nascence of life and exist anywhere except in the Sun and the stars. Therefore, no studies have been practically carried out in this area. The formation of mirror asymmetry was assigned to other factors, such as photochemical processes (Bernstein *et al*., 2002; Munoz Caro *et al*., 2002) (including those in the presence of a magnetic field (Rikken and Raupach, 2000)), seeds of the type of nonsymmetrical quartz crystals (Klabunovskii and Thiemann, 2000), and so on. Nevertheless, if we accept the electrochemical scenario, then apparently the only place in the Solar system where it could possibly be realized is the inner Galilean satellites and, particularly, the satellite which eventually became Amalthea. We readily see that such studies could provide an important clue to the solution of the problem. If future laboratory experiments support the possibility of absolute asymmetric synthesis by electrolysis in the presence of a magnetic field, our scenario will get a strong support. On the other hand, it will become clear that the observed mirror asymmetry of real living substance is only an inevitable consequence of the existence of two different types of charge carriers, electrons and protons, i.e., a consequence of the original barion asymmetry of our Universe.

We should like to point out one more possibility which could favor the nascence and support of life. It appears that nobody has thus far mentioned it. This possibility is again connected with electric current flow. The current not only generates thermal energy; its nonequilibrium electrochemical manifestations can, in principle, serve as an equivalent of photosynthesis or of the use of various exothermal reactions by anaerobic bacteria known to occur in the Earth's conditions. In other words, one could assume the formation of purely "electrical" forms of primitive life, in which the organisms (of lithotropic type) would derive energy not from, say, photosynthesis but from the electric current flowing around and through them.

The existence of forms of life independent of photosynthesis broadens substantially the scope and extent of habitable zones around stars. On the other hand, it appears reasonable to assume that the presence near a star of an Earth-type planet is not a necessary condition for the existence of primitive forms of life. It will suffice if a Jupiter-like giant planet with a system of Galilean satellites is located at a certain distance from the star. Thus we come back in a certain sense to the hypothesis of the Jupiter–Sun being the limiting case of a close binary. It appears to be by now the only planetary cosmogony which is in accord with the fact that a large part of Jovian-like planets discovered to exist near other stars move in high-eccentricity orbits.

Finally, a few words on one more process, inevitable within our scenario, which made possible the transfer of life that appeared in fairly rapidly changing and rough conditions (for life to form, very many versions had to be tried, one after another), not always favorable for its subsequent evolution, to a more peaceful environment, i.e., to the Earth. We are speaking about comets. The idea of their being capable of transporting primitive organisms was put forward more than once; we may recall the theory of panspermia.

Obviously enough, an explosion of the icy envelope of a Galilean satellite must result in a massive meteoroid bombardment of the Earth, which would bring about mass extinction of the biota. Interestingly, each mass extinction was accompanied subsequently by a shift in evolution of the animal kingdom in a new direction. The most impressive example of this is the Cryptozoic/Phanerozoic transition. It is hard therefore to shake off the impression that SP comets, which, as we believe (and what is supported by all available observational manifestations and their chemistry) are the creation and product of the explosions of electrolyzed icy envelopes of distant moonlike bodies, do indeed carry some highly active biological substances and, possibly, primitive organisms with exotic genetic codes.

It is obvious also that life that had formed in the Galilean satellite system was subsequently transported by comets to all bodies of the Solar system and could continue to evolve further, in one form or another. Life found the most favorable conditions for its evolution on the Earth. Here it could develop to the highest forms possessing intellect. The Earth, in its turn, could contribute to the overall process of life evolution on other bodies of the Solar system through hyperimpact ejections from its surface (Drobyshevski, 1995, 2002). Moreover, now it can be predicted that intelligent life will also be transferred in a reasonable future from the Earth to other planets.

The author is indebted to F.T.Robb for a discussion of the possibility of existence of "electrical" forms of life. The work was presented as a talk at the Workshop "Astrobiology Expeditions 2002" (March 25-29, 2002, St.Petersburg, Russia).

REFERENCES

Avalos, M., Babiano, R., Cintas, P., Jimenez, J.L., Palacios, J.C. and Barron, L.D. (1998) Absolute asymmetric synthesis under physical fields: Facts and fictions. *Chem. Revs.*, **98**, 2391-2404.

Avetisov, V.A. and Goldanski, V.I. (1996) Physical aspects of mirror symmetry breaking in the bioorganic world. *Soviet Phys. - Uspekhi*, **166**, 873-891.

Bonner, W.A. (1990) Attempted asymmetric electrochemical reductions in magnetic fields. *Origins Life Evol. Biosphere*, **20**, 1-13.

Bonner, W.A. (1996) The quest for chirality. In "*Physical Origin of Homochirality in Life*", edited by D.B.Cline, AIP Press, pp.17-49.

Bernstein, M.P., Dworkin, J.P., Sandford, S.A., Cooper, G.W. and Allamandola, L.J. (2002) Racemic amino acids from the ultraviolet photolysis of interstellar ice analogues. *Nature*, **416**, 401-403.

Bodenheimer, P., Grossman, A.S., DeCampli, W.M., Marcy, G. and Pollack, J.B. (1980) Calculations of evolution of the giant planets. *Icarus*, 41, 293-308.

Decroly, J.C., Granicher, H. and Jaccard, C. (1957) Caractere de la conductivite electrique de la glace. *Helv. Phys. Acta*, **30**, 465-469.

De Gennes, P.G. (1982) Pierre Curie and the role of symmetry in physical laws. *Ferroelectrics*, **40**, 125-129.

Delahay, P. (1965) *Double Layer and Electrode Kinetics*, Interscience Publishers.

Drobyshevski, E.M. (1974) Was Jupiter the protosun's core? *Nature*, **250**, 35-36.

Drobyshevski, E.M. (1978) The origin of the Solar system: Implications for transneptunian planets and the nature of the long-period comets. *Moon & Planets*, **18**, 145-194.

Drobyshevski, E.M. (1979) Magnetic field of Jupiter and the volcanism and rotation of the Galilean satellites. *Nature*, **282**, 811-813.

Drobyshevski, E.M. (1980) The eruptive evolution of the Galilean satellites: Implications for the ancient magnetic field of Jupiter. *Moon & Planets*, **23**, 483-491.

Drobyshevski, E.M. (1981) The history of Titan, of Saturn's rings and magnetic field, and the nature of short-period comets. *Moon & Planets*, **24**, 13-45.

Drobyshevski, E.M. (1982) On the excess thermal fluxes of Titan and Saturn. *Moon & Planets*, **26**, 33-46.

Drobyshevski, E.M. (1986) The structure of Phaethon and detonation of its icy envelope.

*Earth, Moon, & Planets*, **34**, 213-222.

Drobyshevski, E.M. (1989) Jovian satellite Callisto: Possibility and consequences of its explosion. *Earth, Moon, & Planets*, **44**, 7-23.

Drobyshevski, E.M. (1995) On the hypothesis of hyper-impact ejection of asteroid-size bodies from Earth-type planets. *Intnl. J. Impact Engng.*, **17**, 275-283.

Drobyshevski, E.M. (1997) Post-periJove splitting and lithium overabundance in Shoemaker-Levy 9 favor its planetary origin. *Astron. Astrophys. Trans.*, **13**, 215-224.

Drobyshevski, E.M. (2000) The young long-period comet family of Saturn. *Mon. Not. Roy. Astron. Soc.*, **315**, 517-520.

Drobyshevski, E.M. (2002) Terrestrial origin of SNC meteorites aand shower source for 30 Myr extinctions. *astro-ph*/0204346.

Drobyshevski, E.M., Chesnakov, V.A. and Sinitsyn, V.V. (1995) The electrolytical processes in dirty ices: Implications for origin and chemistry of minor bodies and related objects. *Adv. Space Res.*, **16**, (2)73-(2)84.

Drobyshevski, E.M., Simonenko, V.A., Demyanovski, S.V., Bragin, A.A., Kovalenko, G.V., Shnitko, A.S., Suchkov, V.A. and Vronski, A.V. (1994) New approach to the explosive origin of the asteroid belt. In "*Seventy-Five Years of Hirayama Asteroid Families*", edited by Y.Kozai, R.P.Binzel and T.Hirayama, ASP Conference Series, **63**, 109-115.

Folsom, C.E. (1979) *The Origin of Life*, Freeman & Co, San Francisco.

Hainovski, N.G., Pavluchin, Yu.T. and Hairetdinov, E.E. (1986) Isotopic effect and conductivity mechanism of crystalline $CsHSO_4$. *Solid State Ionics*, **20**, 249-253.

Hobbs, P.V. (1974) *Ice Physics*, Clarendon Press, Oxford.

Kerridge, J.F. (1999) Formation and processing of organics in the early Solar system. *Space Sci. Revs.*, **90**, 275-288.

Klabunovskii, E. and Thiemann, W. (2000) The role of quarz in the origin of optical activity on Earth. *Origins Life Evol. Biosphere*, 30, 431-434.

Miller, S.L. (1953) A production of amino acids under possible primitive Earth conditions. *Science*, **117**, 528-529.

Munoz Caro, G.M., Meierhenrich, U.J., Schutte, W.A., Barbier, B., Segovia, A.A., Rosenbauer, H., Thiemann, W.H.-P., Brack, A. and Greenberg, J.M. (2002) Amino acids from ultraviolet irradiation of interstellar ice analogues. *Nature*, **416**, 403-406.

Ness, N.F., Acuna, M.H., Lepping, R.P., Burlaga, L.F., Behannon, K.W. and Neubauer,

F.M. (1979a) Magnetic field studies at Jupiter by Voyager 1: Preliminary results. *Science*, **204**, 982-987.

Ness, N.F., Acuna, M.H., Lepping, R.P., Burlaga, L.F., Behannon, K.W. and Neubauer, F.M. (1979b) Magnetic field studies at Jupiter by Voyager 2: Preliminary results. *Science*, **206**, 966-972.

Petrenko, V.F. and Chesnakov, V.A. (1990) On a nature of charge carriers in ice. *Soviet Physics - Solid State*, **32**, 2368-2373.

Petrenko, V.F. and Whitworth, R.W. (1999) *Physics of Ice*, Oxford Univ. Press.

Reeves, R.M. (1974) The electrical double layer: The current status of data and models, with particular emphasis on the solvent. In "*Modern Aspects of Electrochemistry*", edited by B.E.Conway and J.O'M.Bockris, **9**, 239-368.

Rhodes, W. and Dougherty, R.C. (1978) Effects of electric and magnetic fields on prochiral chemical reactions: Macroscopic electric and magnetic fields can cause asymmetric synthesis. *J. Amer. Chem. Soc.*, **100**, 6247-6248.

Rikken, G.L.J.A. and Raupach, E. (2000) Enantioselective magnetochiral photochemistry. *Nature*, **465**, 932-935.

Scherban, T. and Nowick, A.S. (1989) Bulk protonic conduction in Yb-doped $SrCeO_3$. *Solid State Ionics*, **35**, 189-194.

Squyres, S.W., Reynolds, R.T., Cassen, P.M. and Peale, S.J. (1983) Liquid water and active resurfacing on Europa. *Nature*, **301**, 225-226.

Stevenson, D.J. and Ashcroft, N.W. (1974) Conduction in fully ionized liquid metals. *Phys. Rev. A*, **9**, 782-789.

Takahashi, F., Tomii, K. and Takahashi, H. (1986) The electrochemical asymmetric reduction of α-keto acids in the magnetic fields. *Electrochim. Acta*, **31**, 127-130.

Washington, J. (2000) The possible role of volcanic aquifers in prebiologic genesis of organic compounds and RNA. *Origins Life Evol. Biosphere*, **30**, 53-79.